\documentclass[aps,pre,nofootinbib,showpacs,eqsecnum,amsmath,
tightenlines]{revtex4}

\usepackage{graphicx}
\usepackage{dcolumn}
\begin{document}

\title{What is the temperature dependence of the Casimir 
effect?}

\author{J. S. H{\o}ye}
\email{johan.hoye@phys.ntnu.no}
\affiliation{Department of Physics, Norwegian University of Science and
Technology, N-7491, Trondheim, Norway}

\author{I. Brevik}
\email{iver.h.brevik@ntnu.no} 
\author{J. B.  Aarseth}
\email{jan.b.aarseth@ntnu.no}
\affiliation{Department of Energy and Process
Engineering, Norwegian University of Science and
Technology, N-7491, Trondheim, Norway}

\author{K. A. Milton}
\email{milton@nhn.ou.edu}
\affiliation{Department of Physics, Washington University, St. Louis,
MO 63130 USA}
\altaffiliation{Permanent address: Homer L. Dodge
Department of Physics and Astronomy, The University of Oklahoma,
Norman, OK 73019 USA}

\date{\today}

\begin{abstract}
There has been recent criticism of our approach to the Casimir
force between real metallic surfaces at finite temperature,
 saying it is in conflict with the third law of thermodynamics and in
contradiction with experiment.  We show that these claims are unwarranted,
and that our approach has strong theoretical support, while the experimental
situation is still unclear.
\end{abstract}


\maketitle

\section{Introduction}

Recently there has been a comment \cite{Bezerra:2005hc}
that has strongly criticized our work \cite{Brevik:2004ue} 
on the temperature dependence of the Casimir force between a
spherical lens and a plate, both coated with real metals (Au for example).
This follows a long paper by some of the same authors \cite{decca},
which also criticizes our work as being in conflict with
fundamental thermodynamical requirements, particularly the third law of
thermodynamics or the Nernst heat theorem.  Furthermore, they claim that
their recent experimental results are completely consistent with their
theoretical approach, and in contrast completely rule out our theory.
The purpose of this note is to respond to these criticisms, and emphasize
that their theoretical claims are invalid, while the experimental situation
is still too indecisive to draw definitive conclusions.

First, let us make a statement about the physical nature of the controversy.
The conventional approach, dating back to Lifshitz \cite{lifshitz80},
and reinforced by Schwinger et al.~\cite{schwinger78}, describes an ideal
metal by a formal $\varepsilon\to\infty$ limit, where $\varepsilon$ is the
permittivity of the material.  Mathematically, the limit is taken in such
a way that both the TE and TM zero modes (that is, the $m=0$ term in
the Matsubara sum) contribute. Recently, however, it has been recognized
\cite{bostrom} that this is not  correct, and that the TE zero mode cannot
contribute either for finite $\varepsilon$ or for a real metal.
Consequently, in the ideal metal limit, this recognition gives
rise to a linear temperature term in the pressure at low temperature,
and a nonzero value of the entropy at zero temperature.  However, real
metals are not described by this ideal limit, and even though
only one zero mode, the TM one,
 can be present, as one sees by either thermodynamic or electrodynamic
 considerations, the contradiction with the Nernst theorem 
 is not present for actual conductors.

In contrast, the authors of \cite{Bezerra:2005hc,decca} propose two 
alternatives
to describe the reflection coefficients that enter into the Lifshitz
theory,  the impedance approach and the plasma model. In the former 
the transverse momentum dependence in the surface impedance is simply 
disregarded (the so-called Leontovich approach), while in the second 
alternative
the plasma dispersion relation with no relaxation is used for the dielectric 
constant. Neither of these omissions is in accordance with the properties
of real materials, as we will detail in the following.

\section{Lifshitz formula}

We start by remarking that the Lifshitz formula 
for the force between two parallel dielectric slabs is essentially geometrical,
being determined entirely by multiple reflections at the 
interfaces between the parallel media (for example, see Sec.~3 of
Ref.~\cite{Milton:2004ya}).
Now there is no doubt that one can use surface impedances to describe the 
reflection coefficients appearing in the theory, for these are merely the 
linear relations between transverse electric and magnetic fields at the 
surface of the metal.  It is well known that use of either bulk permittivity or
surface impedance are equivalent in describing the reflectivity properties,
{\em provided that one incorporates transverse momentum, $k_\bot$.} 
The boundary conditions implied by Maxwell's equations provide the precise 
connection between these alternative descriptions \cite{embook}.  In 
principle both bulk permittivity and surface impedance should depend on 
$k_\bot$, although theory and optical data suggest that the dependence of 
permittivity on $k_\bot$ is rather small.  This would then imply a substantial 
dependence on $k_\bot$ in the surface impedance. Indeed,  in 
Ref.~\cite{Brevik:2003rg} we show that the exact impedance also leads to zero 
contribution of the TE zero mode. 

\subsection{Transverse momentum dependence of the surface impedance}

The authors of Ref.~\cite{Bezerra:2005hc,decca} ignore the transverse
momentum dependence at zero frequency, because they
believe that the ``mass shell'' condition $(\omega/c)^2-k^2_\bot>0$ must be
satisfied.  However, this relation can only be valid for real photons.
In computing the temperature dependence, one is evaluating Green's functions
periodic in imaginary time, with period $\beta=1/kT$ \cite{Martin:1959jp}.
The Matsubara frequencies, $\zeta_m=2\pi m/\beta$, being imaginary, evidently
break the mass-shell condition.  Physically, evanescent waves or virtual
photons are responsible for the thermal Green's functions.  
The inclusion of spatial dispersion, as required by Maxwell's equations, 
has 
been considered in detail in  recent publications \cite{svetovoy,sernelius}
 and the effects of the exclusion of the TE zero mode at both high and low
temperatures are manifested.

\subsection{Plasma model}

The authors of Ref.~\cite{Bezerra:2005hc,decca} also state that the plasma
model is equally good at describing the optical properties of real metals.
The plasma model is a special case of the Drude model, with the relaxation
parameter neglected.  In fact, optical data clearly show that this limit
is inadequate: The Drude model describes the permittivity of real metals up
to $\zeta\approx 2\times 10^{15}$ rad/s, while the plasma model fails for
frequencies below $5\times 10^{13}$ rad/s.  
It is precisely the low-frequency
part of the spectrum that is relevant for the discussion of the disputed
zero-mode contribution.  Those authors argue that the neglect of relaxation
is appropriate because the zero temperature limit of the relaxation 
parameter for an ideal metal (the Bloch-Gr\"uneisen law)
is zero.  To this we reply that the relaxation parameter for
real materials cannot vanish at zero temperature because of the scattering
by impurities, and further that all present and 
nearly all proposed experiments are
carried out at room temperature, so the temperature dependence of this
parameter should be irrelevant.  Moreover, it has recently been demonstrated
\cite{svetovoy} that at sufficiently low temperature the residual value
of the relaxation parameter does not play a role, as the frequency
characteristic of the anomalous skin effect becomes dominant.
In contrast to the ad hoc procedure
advocated in Ref.~\cite{Bezerra:2005hc,decca}, which in addition to
using physically inadequate models, employs an unjustified extrapolation
from the infrared region (that is, the plasma model, which disregards
relaxation, is extrapolated from that region down to zero frequency), 
we use real optical data at room temperature for the bulk permittivity.
Elsewhere it has been emphasized that a dielectric function valid over
a wide frequency range and not merely the infrared must be employed,
and indeed that frequencies very small compared to the characteristic
frequency play a dominant role in the temperature dependence
\cite{svetovoy}. Of course, there is no data at zero frequency, but physical
requirements (thermodynamics and Maxwell's equations) do not allow
the TE zero mode to be included in the Lifshitz formula.

\section{Thermodynamics}

 As far as we can understand there is no disagreement about the validity 
of the Lifshitz formula for the Casimir force as long as the dielectric 
constant
 is finite. For this latter situation there is no $m=0$ or zero frequency TE 
 (transverse electric) mode, in agreement with Maxwell's equations of 
 electrodynamics, and inclusion of such a term would violate the third
law of thermodynamics \cite{svetovoy}.

The controversy arises in the limit of a perfect metal. It is well known that 
in this limit the mathematics of the limiting process can be ambiguous due to 
its singular nature. So the limit of a perfect metal can in one way be taken 
by letting $\varepsilon\rightarrow\infty$ when regarding the contribution from 
the $m=0$ term. Since this term for all finite $\varepsilon$ is zero, its 
limiting value is also zero. However, the other way is to take the limit 
$m\rightarrow 0$ while $\varepsilon=\infty$
\cite{schwinger78}. In the latter case a non-zero 
contribution arises. The opinion of Bezerra et al.~\cite{Bezerra:2005hc} 
is that the latter procedure is the proper one while we have the opposite 
opinion.

So the controversy left is for perfect metals of infinite extension. One of 
our arguments is that this special case is a limiting case of more realistic 
models with relaxation. As mentioned earlier, this latter limit can come in 
conflict with the third law of thermodynamics. But even in this limit the 
possible violation is ambiguous or not obvious. On one hand the entropy 
remains zero for temperature $T=0$ in the limit $\varepsilon\rightarrow\infty$.
On the other hand taking the limit $T\rightarrow 0$ with $\varepsilon=\infty$ 
will result in entropy different from zero. The latter is connected to the 
increasing and diverging slope of the entropy function as 
$\varepsilon\rightarrow\infty$ close to $T=0$. 

However, for realistic models for metals with relaxation (or finite 
conductivity) this controversy is not present. As we have found earlier the
entropy then goes smoothly towards zero as $T\rightarrow 0$. Bezerra et 
al.~\cite{Bezerra:2005hc} correctly point to the fact that the so-called MIM 
(modified ideal metal) violates the third law of thermodynamics. (This we 
stated explicitly in Ref.~\cite{Hoye:2002at}.) This violation is then in the 
sense discussed above, i.e., it is somewhat ambiguous depending upon how the 
limiting process is performed. It is also correct that according to our 
approach real metals approach MIM for increasing separation as then lower 
frequencies become more important. (MIM is the limit where the dielectric 
constant $\varepsilon\rightarrow\infty$ with the TE zero-mode equal to zero.) 
But the crucial difference between real metals and MIM is that the former 
includes relaxation by which there will be no violation of the third law of 
thermodynamics (at least for finite separation of metal plates and for 
infinite separation where there is no force or interaction energy left).
Moreover, numerical studies have shown that the MIM model disagrees strongly at
short distances with values for the Casimir force calculated for real metals 
even for zero temperature.

Bezerra et al.~\cite{Bezerra:2005hc} 
further claim that our unstated assumption is that perfect 
crystals with no defects or impurities do not or cannot exist, and, therefore, 
that thermodynamics can be violated for them. At best this is a 
misinterpretation of our arguments. There is no violation of the 
third law in this connection as quantization of lattice vibrations on a perfect
lattice yields no problem; in fact it can be performed exactly. A reason for 
this is that the standard lattice has a finite dielectric constant and there is
no ambiguity. However, if the perfect lattice turns into a perfect metal 
(with no relaxation) we can again see the problem discussed above. Then there 
will be no well defined thermal equilibrium as any steady current can stay on 
forever. And we do not think that Nernst had this latter very special 
and unrealizable situation in mind when he formulated his theorem. 

Thus, we believe that the results obtained by excluding the TE zero mode
are consistent with the third law of thermodynamics.  The entropy at
zero temperature vanishes, except in the limit of a perfect conductor of
infinite extent, for which thermal equilibrium can never occur.
  There is a region in which the Casimir entropy is negative,
but this is a rather familiar phenomenon, reflecting the fact that we are
describing only part of the complete physical system. 
(Although Svetovoy and Esquivel \cite{svetovoy} and Sernelius \cite{sernelius}
agree with our conclusion that the entropy vanishes at zero temperature
for real metals with no TE zero mode, and thus 
there is no contradiction with the
Nernst theorem, the former authors believe that
 there is a thermodynamic difficulty with the negative entropy region.)

\section{Theoretical support}

Two other recent papers also lend support to our point of view.  Jancovici
and \v Samaj \cite{jancovici} and Buenzli and Martin \cite{buenzli}
have examined the Casimir force between ideal-conductor walls with
emphasis on the high-temperature limit.  Not surprisingly, ideal inert
boundary conditions are shown to be inadequate, and fluctuations within
the walls, modeled by the classical Debye-H\"uckel theory, determine the
high temperature behavior.  The linear in temperature  behavior
of the Casimir force is found to be reduced  by a factor of two
from the behavior predicted by an ideal metal.  This is precisely the
signal of the omission of the $m=0$ TE mode.  Thus, it is very hard to
see how the corresponding modification of the low-temperature behavior
can be avoided. 

Further support for our conclusions can be found in the very recent paper
of Sernelius \cite{sernelius05}, who calculates the van der Waals-Casimir
force between gold plates using the Lindhard or random phase approximation
dielectric function.  Spatial dispersion plays a crucial role in his
calculations.  For large separation, the force is one-half that of the ideal
metal, just as in the calculations in \cite{jancovici,buenzli}.  For arbitrary
separations between the plates, Sernelius' results numerically nearly exactly
coincide with his earlier ones \cite{sernelius01,bostrom} 
 where dissipation (i.e., nonzero relaxation parameter) is
included but no spatial dispersion.  In his new calculation, the inclusion
of dissipation has negligible additional effect.  His new results thus
essentially coincide with ours.

\section{Experimental constraints}

Thus, we claim that there is now overwhelming theoretical evidence
that a TE zero-mode should not be included in the Lifshitz formula when
describing the interaction between real metal surfaces.  This makes the 
purported exclusion of this theory by recent experiments, as forcefully stated
in Refs.~\cite{Bezerra:2005hc, decca}, most difficult to understand.
We believe that the resolution of this conundrum lies in an insufficient
appreciation of the backgrounds making Casimir measurements so difficult.
We and others \cite{iannuzzi} have mentioned the difficulty in
accurately determining the
absolute sphere-plate separation. This may be especially so since the
roughness of the surfaces is much larger than the precision stated in 
the determination of the separation.  Nonlocality in the electric properties
of the surfaces, as for example seen in the calculation of 
Ref.~\cite{jancovici}, means that the location of the surface cannot be
specified to better than some effective shielding length.
Interferometric methods of determining distance may also be somewhat
uncertain because of penetration of the surface by light.
Also accurate determination of a small 
difference between experimental values at room temperature and purely 
theoretical values at $T=0$ gives rise to further difficulties. 
In short, fitting data precisely to a preferred theory wherein various 
unknown experimental parameters must be determined does not constitute
decisive evidence in favor of that theory; nor does the poorness of fit
to an alternative theory when only one of many parameters is allowed to
vary constitute decisive evidence against that theory.

In our view, the issue of temperature dependence cannot be settled until
experiments are able to detect a variation of the Casimir force with 
temperature.  Clearly, such experiments are difficult.  The precise
experiments at very short distances ($\sim 100$ nm), where the Casimir 
forces are largest, are not the best place to look for temperature
variation, for the temperature dependence is relatively small there.  Rather,
experiments should be conducted at the micrometer scale, where the effects,
if our theory is correct, are above the 10\% level, and approach a factor of
two reduction from the ideal metal limit for larger separations. (The Lamoreaux
experiment \cite{lamoreaux}, conducted at the micrometer separation scale, 
was probably not as accurate as claimed.)  Proposals to perform measurements
of the force 
between a cylinder and a plane \cite{brown-hayes} and between eccentric
cylinders \cite{dalvit} have advantages because the forces are stronger
than between a sphere and a plane, yet the difficulties in maintaining
parallelism is not so severe as with two plane surfaces.  In this way
the possiblitity of measuring the temperature dependence at relatively
large distances where thermal effects are largest may be accomplished.

\section{New calculations}

To this end, we present some new calculations of the force between
parallel plates based on our theory, compared with that of 
Ref.~\cite{decca}. 
The numerical computations are done on the basis of Eq. (4.18) in our
paper \cite{Hoye:2002at}, and
follow the recent paper of Bentsen et al.~\cite{Bentsen:2005yi}. 
The maximum value of the quantity $y=
\sqrt{k_\perp^2+\zeta_m^2}a$, $a$ being the plate separation,
 is chosen to be $y_{\rm max}= 30$. In all
calculations we impose the tolerance for the integrals to be $10^{-12}$. 
The overall tolerance in the calculated sum over Matsubara numbers is taken
to be $10^{-8}$. This last tolerance determines the highest Matsubara
frequencies occurring in the $m$ sum. These values are shown in the tables.

As for permittivities, we are using the (new) data received from Astrid
Lambrecht. These data extend
from about $1.5\times 10^{11}$ rad/s to about $1.5\times 10^{18}$ rad/s. 
An important virtue of our tables is that they show which frequency domains we 
actually use. The largest frequency region naturally occurs when the 
temperature is low. Thus for $T=1$ K, the temperature that we associate with 
$T=0$ on physical grounds, we use the region from about $8\times10^{11}$ rad/s 
to about $2\times10^{16}$ rad/s. That means, we are, even at this low 
temperature, working with frequencies that lie entirely within the region of 
Lambrecht's data. This implies that we do not have to involve the Drude 
relation at all, for the finite frequencies. Of course, the Kramers-Kronig
relation, required by causality, is employed. There is no analytical 
approximation involved in our formalism, for any finite frequency. The only 
exception is the zero frequency case. We then need the Drude relation to 
assure that there is no contribution to the force from the zero frequency TE 
mode. This zero frequency contribution to the Casimir force is found 
analytically, not numerically.

For comparison, we show in Table \ref{tab1} the results given in
Ref.~\cite{decca} for the Leontovich impedance and plasma models.
These results are for zero temperature.  In their approach the temperature
dependence is negligible.  It will be noted that their two models
do not agree, even though the authors seem to imply that either model
is equally good.  Our view is that the difference between these two
models may be taken as a rough gauge of the accuracy of their predictions.
Our results, with details about the frequency
range used and the number of terms in the Matsubara sum, are given in
Tables \ref{tab2}--\ref{tab4} for $T=1$ K (sufficiently close to $T=0$ K,
but a temperature at which our numerical technique is stable), $T=300$ K,
and $T=350$ K.  The latter is given in the hope that the temperature
variation over a 50 K range near room temperature
 may soon become accessible to experiment.

\begin{table}
\begin{tabular}{ccc}
\toprule
Separation&Impedance Method&Plasma Method\\
nm&Ref.~\cite{decca}&Ref.~\cite{decca}\\
\colrule
160&1144&1114.9\\
200&509.3&501.8\\
250&224.7&223.1\\
400&38.90&38.98\\
500&16.70&16.76\\
700&4.605&4.628\\
\botrule
\end{tabular}
\caption{\label{tab1} Casimir pressure between parallel plates in the two 
models discussed in Ref.~\cite{decca}.
Pressures are given in mPa.}
\end{table}

\begin{table}
\begin{tabular}{cccc}
\toprule
Separation&Pressure&Highest frequency&Number of terms\\
nm&mPa&rad/s&in sum\\
\colrule
160&1144&$2.112\times10^{16}$&25674\\
200&508.2&$1.713\times10^{16}$&20824\\
250&223.7&$1.388\times10^{16}$&16869\\
400&38.61&$8.875\times10^{15}$&10789\\
500&16.56&$7.168\times10^{15}$&8714\\
700&4.556&$5.187\times10^{15}$&6305\\
1000&1.143&$3.674\times10^{15}$&4466\\
\botrule
\end{tabular}
\caption{\label{tab2} Our results for the Casimir pressure between gold
plates, when $T= 1$ K.  The first nonvanishing Matsubara frequency
(corresponding to $m=1$) is  $8.226\times 10^{11}$ rad/s. 
The highest frequencies are
dependent on the plate separation $a$,  
as shown, and are determined by our chosen  tolerance 
$10^{-8}$ for the sum. The last column gives the number of terms in the sum.}
\end{table}

\begin{table}
\begin{tabular}{cccc}
\toprule
Separation&Pressure&Highest frequency&Number of terms\\
nm&mPa&rad/s&in sum\\
\colrule
160&1127&$2.122\times10^{16}$&86\\
200&497.8&$1.727\times10^{16}$&70\\
250&217.6&$1.407\times10^{16}$&57\\
400&36.70&$8.884\times10^{15}$&36\\
500&15.49&$7.403\times10^{15}$&30\\
700&4.127&$5.429\times10^{15}$&22\\
1000&0.9852&$3.702\times10^{15}$&15\\
\botrule
\end{tabular}
\caption{\label{tab3} Same as in Table \ref{tab2}, but at $T=300$ K. The first
nonvanishing Matsubara frequency is $2.468\times 10^{14}$ rad/s. The highest
frequencies are determined by the same tolerance $10^{-8}$ for the $m$ sum as
before.}
\end{table}

\begin{table}
\begin{tabular}{cccc}
\toprule
Separation&Pressure&Highest frequency&Number of terms\\
nm&mPa&rad/s&in sum\\
\colrule
160&1124&$2.131\times10^{16}$&74\\
200&495.7&$1.727\times10^{16}$&60\\
250&216.4&$1.411\times10^{16}$&49\\
400&36.35&$8.925\times10^{15}$&31\\
500&15.30&$7.198\times10^{15}$&25\\
700&4.052&$5.470\times10^{15}$&19\\
1000&0.9590&$3.743\times10^{15}$&13\\
\botrule
\end{tabular}
\caption{\label{tab4} Same as in Table \ref{tab3}, but at $T=350$ K. The first
nonvanishing Matsubara frequency is $2.879\times 10^{14}$ rad/s. The highest
frequencies are determined by the same tolerance $10^{-8}$ for the $m$ sum as
before.}
\end{table}

\section{Conclusions}
Were the controversy over the temperature dependence a purely theoretical
one, we believe that a consensus along the lines we have been advocating,
justified by  realistic model calculations by Jancovici and \v Samaj 
\cite{jancovici}, Buenzli and Martin \cite{buenzli}, and Sernelius
\cite{sernelius05}, would now be nearly universally accepted.  Evidently
what has prevented this consensus from developing is the apparent conflict
that the resulting predictions have with experiment.  However, we believe
it is premature to regard that conflict as decisive.  
It is imperative to perform room-temperature experiments at larger separations,
since the relative temperature corrections are then much larger. The optimum 
separation seems to lie around 2 $\mu$m \cite{ba}. Indeed, the early
experiments of Lamoreaux \cite{lamoreaux} support the Drude theory at
large separation, although the error bars are large. The experimental situation
when $a$ is about 1 $\mu$m or less (which is the main region of the Decca 
experiment \cite{decca}) may be influenced by extra factors coming into play 
at short distances, such as surface plasma effects \cite{il}.
At short distances, the results are moreover very sensitive to differences in 
separation; an error of 1 nm may give rise to errors in the pressure of about 
10 mPa or 2\% at $a=200$ nm.

\acknowledgments{We thank V. M. Mostepanenko and G. L. Klimchitskaya for
alerting us to some numerical discrepancies in the data we had used
in an earlier version of this paper. We also thank the other
participants of the Seventh
Workshop on Quantum Field Theory Under the Influence
of External Conditions in Barcelona for numerous helpful comments.
We thank Astrid Lambrecht for
supplying us with new data.  We are grateful to R. Cowsik for
conversations on the difficulty of Casimir force measurements.  
And we thank Bo Sernelius for making his preprint \cite{sernelius05}
available to us. The work of
K. A. M. is supported in part by grants from the US Department of Energy.
He also thanks the Department of Physics of Washington University for its
hospitality and support.}

\end{document}